\documentclass[%
 reprint,
nofootinbib,
 amsmath,amssymb,
 aps,
]{revtex4-1}

\usepackage{graphicx}
\usepackage{dcolumn}
\usepackage{bm}

\usepackage{color}
\usepackage{here}

\begin{document}


\title{Spin--gap formation due to spin--Peierls instability in $\pi$--orbital--ordered NaO$_2$}

\author{Mizuki Miyajima$^{1,\S, *}$, 
Fahmi Astuti$^{2,3,\P, *}$, 
Takahito Fukuda$^1$, 
Masashi Kodani$^1$, \\
Shinsuke Iida$^4$, 
Shinichiro Asai$^4$, 
Akira Matsuo$^4$, 
Takatsugu Masuda$^{4,6,7}$, 
Koichi Kindo$^4$, \\
Takumi Hasegawa$^5$, 
Tatsuo C Kobayashi$^{1}$, 
Takehito Nakano$^8$, 
Isao Watanabe$^{2,3}$, and 
Takashi Kambe$^{1,\dag}$ \email{kambe@science.okayama-u.ac.jp}}

\affiliation{%
$^1$Department of Physics, Okayama University, Okayama 700--8530, Japan
}%
\affiliation{%
$^2$Advanced Meson Science Laboratory, RIKEN Nishina Center, Wako, Saitama 351--0198, Japan
}%
\affiliation{%
$^3$Department of Physics, Hokkaido University, Sapporo 060--0808, Japan
}%
\affiliation{%
$^4$Institute for Solid State Physics, The University of Tokyo, Kashiwa, Chiba 277--8581, Japan
}%
\affiliation{
$^5$Graduate School of Advanced Science and Engineering, Hiroshima University, Higashi--Hiroshima, 739--8521, Japan
 }
\affiliation{
$^6$ Institute of Materials Structure Science, High Energy Accelerator Research Organization, Tsukuba, Ibaraki 305--0801, Japan
}
\affiliation{
$^7$ Trans--scale Quantum Science Institute, The University of Tokyo, Tokyo 113--0033, Japan 
}
\affiliation{
$^8$Institute of Quantum Beam Science, Ibaraki University, Mito, Ibaraki 310--8512, Japan
}

\date{\today}

\begin{abstract}
We have investigated the low--temperature magnetism of sodium superoxide (NaO$_{2}$), in which 
spin, orbital, and lattice degrees of freedom are closely entangled. 
The magnetic susceptibility shows anomalies at $T_{1}=220$ K and $T_{2}=190$ K, 
which correspond well to the structural phase transition temperatures, and a sudden decrease below $T_{3}=34$ K. 
At 4.2 K, the magnetization shows a clear stepwise anomaly around 30 T with a large hysteresis. 
In addition, the muon spin relaxation experiments indicate no magnetic phase transition down to $T=0.3$ K. 
The inelastic neutron scattering spectrum exhibits magnetic excitation with a finite energy gap. 
These results confirm that the ground state of NaO$_{2}$ is a spin--singlet state. 
To understand this ground state in NaO$_{2}$, we performed Raman scattering experiments. 
All the Raman--active libration modes expected for the marcasite phase below $T_{2}$ are observed. 
Furthermore, we find that several new peaks appear below $T_{3}$. 
This directly evidences the low crystal symmetry, namely, the presence of the phase transition at $T_{3}$. 
We conclude the singlet--ground state of NaO$_{2}$ due to the spin--Peierls instability. 
\end{abstract}

\pacs{Valid PACS appear here}
\maketitle


The entanglement of spins, orbitals, charge, and lattice degrees of freedom is a fundamental issue in solid--state physics \cite{Tokura, Goodenough, Kugel, Kanamori}. 
They yield a variety of phenomena such as superconductivity, quantum liquids, multi--ferroics, and orbital liquids 
\cite{BaCuSbO, Kimura, Ishihara}.  
While these physics have been discussed mainly in $d$-- and $f$--electron systems, 
there has been very little discussion in $p$-- or $\pi$--electron systems \cite{fullerene, tdae_kambe}. 
The magnetism of alkali--metal superoxide,  $A$O$_2$, originates from the unpaired $p$--electrons on O$_2$ molecule. 
In O$_{2}$, as two unpaired electrons occupy two $\pi _{g}^*$--orbitals, $\pi_{x}^{*}$ and $\pi_{y}^{*}$, 
whose spins are in parallel to each other, 
O$_2$ is a magnetic molecule with the spin quantum number, $S$, of 1. 
In $A$O$_2$, as $A$--cation is usually fully ionized to form $+1$ state, 
the additional electron on O$_{2}$ occupies one of two half--filled $\pi _{g}^*$--orbitals. 
This allows the O$_2^{-}$ molecule to gain the orbital degrees of freedom. 
As three electrons exist on two $\pi _{g}^*$s, the energy band for the O$_2^{-}$ state is quarter--filled.   
The {\it ab initio} band calculations for $A$O$_{2}$ show that the Fermi energy locates within the $\pi$--band \cite{KO2_kim, RbO2_Mott, KO2_Solovyev}. 
On the contrary, the experimental magnetic properties suggest electron localization, which implies the importance of an 
electronic correlation on the O$_{2}$ molecule. 
Thus, $A$O$_{2}$ should be considered as a Mott insulator \cite{KO2_kim, RbO2_Mott}. 
Spontaneous Jahn--Teller distortion should occur to lift the degeneracy of $\pi_{g}^*$--orbital, 
which leads to the selection of the $\pi^*$--orbitals. 
The coherent arrangement of the O$_2$ orbitals is expected to lead to a three--dimensional magnetic exchange interaction between the spins. 
Therefore, in $A$O$_2$, spins, orbitals, charge and lattice degrees of freedom are strongly coupled \cite{KO2_Solovyev}. 
$A$O$_{2}$ is expected to be a candidate material to exhibit such fascinating phenomena. 

Recently, $A$O$_2$ has attracted much attention for its magnetic quantum phenomena 
at low temperatures \cite{CsO2_Riyadi, CsO2_NMR, CsO2_ESR, CsO2_Miyajima, RbO2_Fahmi, Fahmi_thesis, Miyajima_thesis}. 
CsO$_2$ shows one--dimensional (1D) antiferromagnetic (AF) behavior in the magnetic susceptibility 
and the high--magnetic field magnetization \cite{CsO2_Riyadi, CsO2_Miyajima}. 
It was suggested that a 1D chain should form as a result of the $\pi$--orbital ordering, 
but the detailed low--temperature structure has not been determined \cite{CsO2_Riyadi}. 
NMR experiments showed a power--law dependence of the spin--lattice relaxation function, 
suggesting the emergence of a Tomonaga--Luttinger Liquid state in the 1D short range ordered phase \cite{CsO2_NMR}. 

Contrary to RbO$_2$ and CsO$_2$, 
NaO$_{2}$ has a cubic (Space group; $Fm$\={3}$m$) symmetry at room--temperature, 
in which O$_2$ has an orientational disorder \cite{AO2_XRD}. 
With decreasing temperature, NaO$_2$ shows successive structural phase transitions 
at $T_1=220$ and $T_2 \sim 196$ K \cite{AO2_XRD}. 
In the marcasite phase below $T_2$, the degeneracy of the $\pi_{g}^{*}$ orbitals is considered to be lifted due to the low local symmetry around O$_{2}$. 
The magnetic susceptibility shows a weak decrease below $T_{2}$ and, then, a sudden drop below $T_{3} =30\sim 40$ K. 
These experimental findings remind us of a low dimensionality of the spin system and 
of a magnetic phase change below $T_{3}$. 
Theoretical calculations for the marcasite phase pointed to the quasi--1D AF character along the $c$--axis 
and frustration of exchange interactions between different sublattice spins \cite{NaO2_Solovyev}. 
However, because of no experimental inspection on an existence of magnetic phase transition and a change in crystal symmetry, 
the magnetic ground state of NaO$_{2}$ has not yet been clarified. 
In this Letter, we investigate low--temperature magnetic and structural properties of the $\pi$--orbital system NaO$_{2}$ in detail. 
For this purpose, we performed magnetic susceptibility, high--field magnetization, muon spin relaxation ($\mu$SR), 
x--ray diffraction (xrd), inelastic neutron scattering (ins) and Raman scattering experiments using high quality samples. 

\begin{figure}[t]
\begin{center}
\includegraphics[clip, width= 0.45\textwidth]{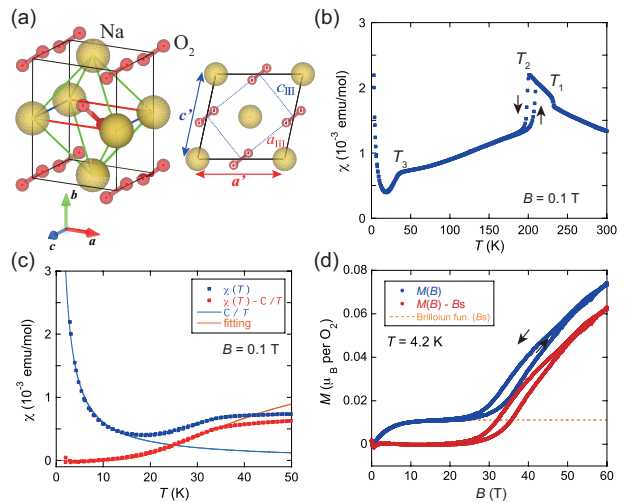}
\caption{
(a) Unit cell of the marcasite phase of NaO$_{2}$. 
The right figure shows the unit cell in the ($ac$)--plane. 
$a_{\rm III}$ and $c_{\rm III}$ denote the axes in the phase III while $a'$ and $c'$ correspond to that in the phase II. 
(b) Temperature dependence of magnetic susceptibility, $\chi (T)$, at $B=0.1$ T in powder NaO$_2$. 
$\chi (T)$ shows clear anomaly at $T_{1}$, $T_{2}$ and $T_{3}$. 
The arrows indicate the hysteresis in the cooling and heating protocols. 
(c) Enlarged figure around $T_3$. 
The experimental data $\chi (T)$ (blue dot), the low--temperature Curie--tail $C_{0}/T$ (blue line) and the subtracted 
data $\chi (T)-C_{0}/T$ (red dot) are shown. 
$\chi(T)$ is fitted by the equation in the text. 
(d) High--magnetic field magnetization at 4.2 K, where the vertical axis indicates the magnetization per O$_2$. 
At low--magnetic field region, the $M(B)$ curve can be fitted by the Brillouin function, $B_{s}$. 
The experimental data $M (B)$ (blue dot), the $B_{s}$ (orange dotted line) and the subtracted 
data $M (B)-B_{s}$ (red dot) are shown.
}
\label{magnetization}
\end{center}
\end{figure}

First, we define three phases \cite{AO2_XRD}; the phase I above $T_{1}$, 
the phase II between $T_{1}$ and $T_{2}$, the phase III between $T_{2}$ and $T_{3}$. 
NaO$_2$ has a remarkable temperature dependence of magnetic susceptibility, 
$\chi = M/B$, where $M$ and $B$ denote the magnetization and the magnetic field, respectively. 
Figure \ref{magnetization} (b) shows the temperature dependence of $\chi (T)$ for a powder sample 
using a cooling and heating protocol with a magnetic field of 0.1 T. 
$\chi (T)$ shows anomalies around $T_1$, $T_2$ and $T_3$, 
which is consistent with the literature \cite{AO2_magnetism}. 
The temperatures of $T_1$ and $T_2$ correspond well to the structural phase transition temperatures \cite{supple_1}. 
In the phase I, the $\chi (T)$ follows the Curie--Weiss law with a negative Weiss constant of $\theta = -9.4$ K. 
Small $\theta$ indicates a weak AF interaction, which should be due to the orientational disorder of O$_2$. 
In the phase II, the $\theta$ changes to $41.1$ K, suggesting ferromagnetic correlations due to 
the orientational ordering of O$_2$. 
The effective magnetic moments above and below $T_1$ are estimated to be about 1.82 and 1.68 $\mu_{\rm{B}}$, respectively, which increase slightly from the value expected from the spin only. 
This may be due to the orbital effect. 
Following a clear hysteresis around $T_2$, 
which indicates a first order phase transition, $\chi (T)$ shows a weak decrease with decreasing temperature. 
As the localized spins on O$_2$ are responsible for the magnetism of NaO$_2$, 
the remarkable decrease of $\chi (T)$ in the phase III is an important key to consider 
the low--temperature magnetism \cite{NaO2_Mahanti, NaO2_Solovyev}. 
In other words, it may be reasonable to think that it comes from low--dimensionality. 
To evaluate $J$ from the $\chi (T)$, we used the Bonner--Fisher model \cite{BF} and 
a two--dimensional model with weak inter--chain interaction \cite{Keith}, as shown in Fig. S3 in Supplemental Material. 
However, we could not reproduce the experiments by these models using the temperature--independent $J$. 
The xrd experiments showed that the thermal shrinkage coefficient of the $c$--axis was 
$-5.16\times 10^{-4}$ (\AA/K), which was the largest value among the principal axes. 
Thus, the shrinkage of the nearest neighbor (NN) length between O$_{2}$s, which corresponds to the $c$--axis length, 
may be sufficient to make AF $J$ stronger as the temperature decreases. 

Below $T_3$, $\chi (T)$ decreases rapidly with decreasing temperature with no temperature hysteresis 
and, then, increases. 
The $\mu$SR experiment showed that no magnetic ordering was found down to 0.3 K 
as described later. 
Therefore, the decrease of $\chi (T)$ implies the appearance of a spin--gap in the spin excitation spectrum. 
Note that the low--temperature Curie--tail is strongly dependent on the sample batch, 
indicating that it is due to the extrinsic spins. 
To evaluate the intrinsic temperature dependence of $\chi (T)$, we use the equation:  
$\chi (T)= C_{0}/T + C/T \exp\left({ - 2\Delta/k_{\rm B}T}\right)$
, where the $\Delta$ denotes the spin--gap and $k_{\rm B}$ is the Boltzmann constant \cite{muSR_NaTiSiO, SP_NMR}. 
The first term is the Curie--tail contribution, which can be subtracted from the data, and 
the second term is responsible for the spins excited from the singlet state to the magnetic excited states. 
As shown in Fig. \ref{magnetization} (c), the fitting by this equation 
with $\Delta / k_{\rm B} =51.2$ K is good. 
The Curie tail allows us to estimate the number of extrinsic spins with $S=1/2$ to be 0.014 mol in this sample. 

Figure \ref{magnetization} (d) shows the magnetization curve, $M(B)$, as a function of $B$ up to 60 T at $T=4.2$ K.  
The saturation magnetization is equivalent to $\sim 1$ $\mu_{\rm B}$. 
Note that the $M(B)$ experiment was performed using the same sample as the $\chi (T)$ experiments. 
At $T=4.2$ K, the $M(B)$ shows a nonlinear increase at low fields and an anomaly with a large hysteresis around $\sim 30$ T. 
To evaluate the low--magnetic field part of the $M(B)$ curve, we use a Brillouin function as shown in Fig. \ref{magnetization} (d). 
The fitting was good and the number of paramagnetic spins was obtained to be 0.011 mol. 
This value is consistent with that obtained from the Curie--tail in the $\chi (T)$ experiment, 
allowing us to subtract the low--field Brillouin contribution from the $M(B)$ curve. 
Accordingly, $M(B)$ shows the magnetic field induced transition from non--magnetic 
to magnetic state with the large hysteresis around $\sim 30$ T. 

\begin{figure}[t]
\begin{center}
\includegraphics[clip, width = 0.45 \textwidth]{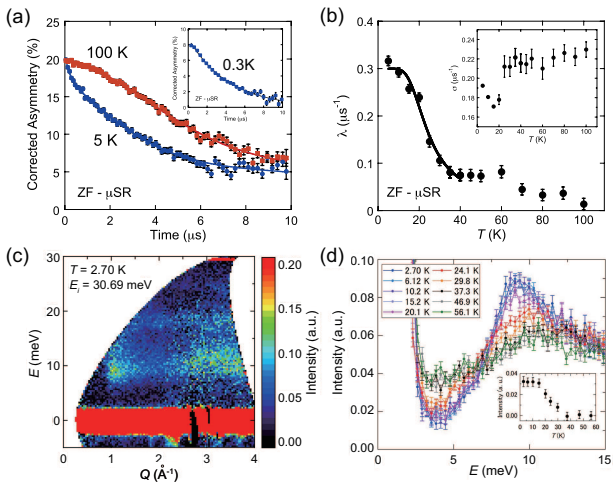}
\caption{
(a) Zero--field (ZF) $\mu$SR spectra of NaO$_{2}$ at 5 K and 100 K. 
The inset shows ZF $\mu$SR spectrum at 0.3 K.
The solid lines are the fitted curves using the equation in the text. 
(b) Temperature dependence of $\lambda$. 
The solid line is the fitted curve using the gap--related analysis function described in the text \cite{muSR_NaTiSiO, SP_NMR}. 
The inset shows the temperature dependence of $\sigma$. 
(c) ins spectra after background subtraction. 
Details are described in Supplemental Material. 
(d) ins profiles, where the spectra between 0.5 \AA $^{-1}$ and 1.5 \AA $^{-1}$ are integrated. 
Temperature dependence of the intensity at 9.4 meV is shown in the inset. 
}
\label{muSR_neutron}
\end{center}
\end{figure}

To find out any signatures of a magnetic ordering in NaO$_2$, we performed $\mu$SR experiments down to 0.3 K. 
Figure \ref{muSR_neutron}(a) shows corrected time spectra measured at 0.3 K ($\ll T_3$), 5 K and 100 K ($\gg T_3$) 
in the zero--field (ZF) condition. 
Neither the loss of the initial asymmetry at $t$ = 0 nor the muon--spin precession was observed down to 0.3 K. 
These findings exclude the presence of a long--range magnetic order down to 0.3 K. 

To focus on collecting some data points around $T_3$, we analyzed the depolarization rate obtained in ZF 
by using the function: $A(t) = A e^{-\lambda t} e^{-(\sigma t)^2} + {\rm B.G.}$ \cite{Kubo_Toyabe}. 
In here, the Gaussian term arises from the muon--spin relaxation caused by randomly distributed internal fields coming from surrounding nuclear dipoles. 
The exponential term ascribes the effect of the fluctuating electronic moments around the muon\cite{Muon_Relax_1,Muon_Relax_2}. 
The B.G. indicates background signals from muons which do not stop in the sample but in a sample mounting plate. 
These B.G. signals were subtracted from the raw signal as a constant term in order to achieve corrected $\mu$SR time spectra. 

The temperature dependence of $\lambda$ and $\sigma$ are shown in Fig. \ref{muSR_neutron} (b). 
Above $T_3$, we obtained $\lambda \sim 0.1$ and $\sigma \sim 0.22$ $\mu$s$^{-1}$. 
The result of $\lambda$ obtained above $T_3$ indicates that electronic spins likely fluctuate beyond the $\mu$SR characteristic time window ($10^{-6} \sim 10^{-11}$ sec), resulting in the motional narrowing limit. 
The $\lambda$ increases below $T_3$ suggests that the muon is expected to sense the formation of the spin--gap state 
in NaO$_2$. 
One possible scenario to explain this result is that the muon spin relaxes its polarization by the thermally activated electronic spins across the spin--gap. 
The same behavior was observed in other spin--gap systems \cite{muSR_NaTiSiO}. 
Following this scenario, the spin--gap is estimated from the temperature dependence of $\lambda$ below $T_3$ 
by applying the following function: $\lambda (T) =\lambda_0 \left\{ 1+ C' \exp(- 2 \Delta/k_{\rm{B}}T) \right\}^{-1}$ \cite{muSR_NaTiSiO, SP_NMR}. 
Using this equation, we estimated the $\Delta/k_{\rm{B}}$ to be $\sim$ 44.6 K, 
which was consistent with that obtained from the magnetic susceptibility measurement. 

Figure \ref{muSR_neutron}(c) shows the ins spectra of powder sample at 2.70 K, 
where the background contribution was subtracted (see Supplemental Material). 
The excitation around $Q \sim 1$ \AA$^{-1}$ is shown to have a finite energy gap. 
The intensity decreases with $Q$, which is typical behavior of magnetic scattering, in the range of $Q \lesssim 2$ \AA $^{-1}$. 
The enhanced intensity at $Q \gtrsim 2$ \AA $^{-1}$ is from remnant phonon scattering of the sample cell made of Aluminum.  
The first momentum of the dynamical structure factor in 1D AF spin chain is proportional to 
$1-\sin Qd/Qd$, where $d$ is the distance between spins~\cite{Igor}, leading to 
the pronounced intensity at $Q \sim 1$ \AA$^{-1}$. 

To reveal the change in the intensity as a function of temperature, 
the spectra integrated between 0.5 \AA $^{-1}$ and 1.5 \AA $^{-1}$ are shown in Fig. \ref{muSR_neutron}(d).  
At 2.70 K, the intensity starts to increase at $E \sim 4$ meV and has a maximum at $E \sim 9$ meV. 
The temperature dependence of the intensity at the maximum energy is shown in the inset. 
It shows no temperature dependence above $T_{3}$, and gradually increases with decreasing temperature below $T_{3}$. 
This result directly indicates that NaO$_2$ has the magnetic excitation with an excitation gap energy of 9 meV below $T_{3}$. 
As the $\mu$SR experiments indicated no magnetic long--range ordering down to 0.3 K, 
this peak results not from magnetic excitation in the magnetic long--range ordered phase but 
from singlet--triplet excitation in the non--magnetic ground state. 

\begin{figure}[h]
\begin{center}
\includegraphics[clip, width= 0.45\textwidth]{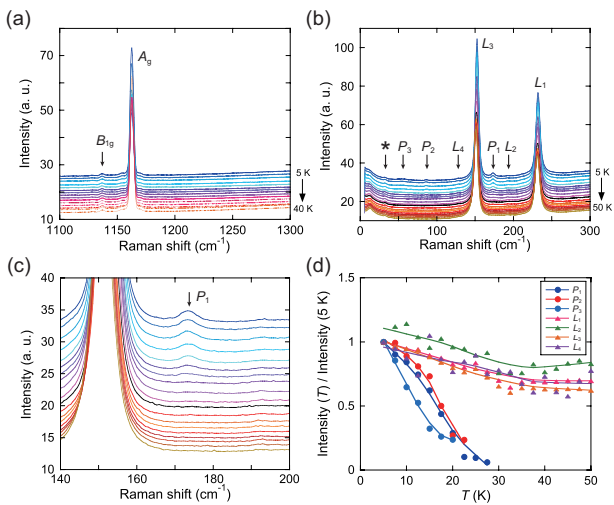}
\caption{Raman scattering results of NaO$_2$. 
(a) and (b) 
Temperature dependences of the stretching modes and the libration modes 
below 50 K, respectively. 
The spectra are shifted along the vertical axis for clarity. 
The asterisk peak is a line included in the laser source. 
(c) Enlarged figure around 150 cm$^{-1}$ for the libration modes. 
(d) Temperature dependences of peak intensity for the libration mode region, 
where the peak intensities are normalized by the intensity at 5 K. 
Solid lines are guides to the eye. }
\label{raman}
\end{center}
\end{figure}

We searched for structural dimerization of the O$_{2}$ molecules as the cause of the non--magnetic state 
by the xrd and neutron diffraction measurements, 
but could not experience the direct evidence on the structural change below $T_{3}$ \cite{supple_2}. 
Then, we perform Raman scattering experiments because of high sensitivity to changes in crystal symmetry and/or molecular charge. 
Figures \ref{raman} (a) and (b) 
show the temperature dependence of the Raman scattering spectra in 
the stretching and the libration mode region of O$_{2}$, respectively. 
We will focus on the change of the Raman--active modes around $T_3$. 

As the crystal symmetry of the phase III is determined as $Pnnm$ ($D_{2h}$), 
the sets of Raman active stretching and libration modes are given by $\Gamma_S = A_{g} + B_{1g}$ 
and $\Gamma_L = B_{1g} +B_{3g} + A_{g} + B_{2g}$, respectively \cite{AO2_Raman, AO2_opt}. 
As shown in Fig. \ref{raman} (a), two peaks can be clearly observed in the stretching mode region. 
We can assign that the peaks at 1163 cm$^{-1}$ and 1140 cm$^{-1}$ originate from the in--phase and the out--of--phase stretching modes, namely, $A_{g}$ and $B_{1g}$, respectively. 
The $B_{1g}$ mode is observed for the first time. 
In the libration mode region (see Fig. \ref{raman} (b)), 
two major peaks are observed around 150 cm$^{-1}$ ($L_{3}$) and 240 cm$^{-1}$ ($L_{1}$).  
Moreover, very weak peaks around 130 cm$^{-1}$ ($L_{4}$) and 190 cm$^{-1}$ ($L_{2}$) are observed for the first time. 
These libration modes can be assigned from the intensity and energy, but a more detailed consideration will be needed. 
Anyway, these results demonstrate that all Raman active modes for the phase with the $D_{2h}$ symmetry can be successfully detected \cite{AO2_Raman}. 
Below $T_3$, while no change was observed in the stretching mode region, 
in the libration mode region, new peaks at 173, 86 and 56 cm$^{-1}$ gradually appeared, which are represented as $P_1$, $P_2$ and $P_3$ in Fig. \ref{raman} (b), respectively. 
Figure \ref{raman} (c) shows the temperature dependence of the $P_{1}$ peak and 
Fig. \ref{raman} (d) summarizes the temperature dependence of the peak intensity observed in the libration mode region. 
While the $\Gamma_L$ modes depended weakly on the temperatures, 
the $P_1$, $P_2$ and $P_3$ peak intensities increased markedly below 30 K, following order parameter like behavior. 
Note that no splittings of the stretching modes are found, suggesting an absence of charge ordering on O$_{2}$. 
Thus, this result clearly indicates an existence of a phase transition around $T_3$. 
Because all Raman--active $\Gamma_L$ modes are confirmed in the phase III, 
the observation of the new peaks is direct evidence for the low crystal symmetry below $T_3$. 
%

To discuss the cause of  the non--magnetic state, 
it is necessary to understand the crystal structure of the precursor phase, i.e., the phase III. 
O$_{2}$ is octahedrally surrounded by Na atoms in all phases. 
In the phase II, the molecular axis is aligned along one of the four equivalent [111]--directions of the 
octahedron (see Supplemental Material). 
The NN molecules arrange their molecular axes to avoid each other, 
i.e., coherent {\it antiferro}--like arrangement of the molecular axes. 
Because the symmetry of the octahedron is still cubic, the degeneracy of the $\pi_{g}^{*}$ orbital should 
be conserved. 
On the contrary, in the phase III, the three--fold symmetry of the octahedron is lost and the molecular axis is slightly 
tilted from the [111]--direction of the octahedron. 
Na--Na bond lengths are changed to be not equivalent and the Na--O bond length is increased along the $b$--axis 
(In Fig.\ref{magnetization} (a), different Na--Na bond lengths are displayed by colors). 
The NN molecules within the $ac$--plane arrange their molecular axes to be parallel to each other, 
namely, the {\it ferro}--like arrangement of the molecular axes is realized. 
Thus, the NN molecular axis is parallel to each other along the $c$--axis 
while those is canted along the $a$-- and $b$--axis. 
The two--fold distortion of the octahedron should break the orbital degeneracy, and stabilizes 
the $\pi_{g}^{*}$ orbital perpendicular to the $c$--axis as unoccupied orbital. 
Namely, a {\it ferro}--orbital ordering realizes in the $ac$--plane. 
The $c$--axis length, i.e., the length between the NN molecules, is obtained to be 3.39 \AA$ $ at 100 K, 
which is close to the length between molecules in the $\alpha$--phase of the solid--O$_2$ ($\sim 3.2$ \AA) \cite{O2_Defoits, O2_Date}. 
Consequently, we can expect the strong AF exchange interaction along the $c$--axis. 
This structural peculiarity should be manifested in the $\chi(T)$ below $T_{2}$, namely, the low--dimensional nature. 
Moreover, 
because there was no structural dimerization of O$_{2}$ in any directions in all phases, 
we were able to deny both isolated dimerization of O$_{2}$ and 1D alternating AF chain as the cause of the spin--gap.  
Accordingly, we can conclude that the magnetism in the phase III is based on the uniform AF spin--chain and, 
then, the magnetic ground state is the spin--Peierls (SP) state. 

Finally, we consider the SP state in NaO$_{2}$. 
As the spin--gap value and the SP transition temperature are obtained to be $\Delta/k_{\rm B} = 51.2$ K and 
$T_{3}=T_{\rm SP}=34$ K, respectively, from the magnetic measurements, 
the value of $2\Delta/k_{\rm B} T_{\rm {SP}} $ is calculated to be 3.01. 
This is comparable with the BCS weak coupling result (3.54). 
The alternating exchange interaction constants, $J_{1}$ and $J_{2}$, in the SP state can be written 
as $J_{1,2}=J(1 \pm \delta)$ using the alternating parameter $\delta$ \cite{SP_Pytte}.  
The spin--gap is also related to $\Delta = 2 p J \delta$ with $p \sim 1+ 2/\pi$. 
Using the Bulaevskii's formula for the $\chi (T)$ below the SP transition \cite{Bulaevskii}, 
we estimate the $\delta$ by fitting to the experiment. 
When we use the $J/k_{\rm B}=140$ K in the 1D uniform AF phase, which was obtained around 50 K, 
we obtain the $\delta$ of 0.11. 
For the organic SP compounds TTF--CuBDT \cite{TTF_CuBDT_Bray} and the inorganic CuGeO$_{3}$ \cite{CuGeO3_Hase}, 
the $\delta$ was estimated to be 0.167 and 0.167, respectively. 
The $\delta$ in NaO$_{2}$ is comparable to these values. 
Moreover, the $\delta$ of 0.11 roughly leads to the $\Delta/k_{\rm B}$ of 51 K, which is almost identical to the spin--gap value obtained in the experiment. 
It is known that the AF $J$ of the solid--O$_{2}$ magnet depends exponentially on the inter--molecular length \cite{O2_Hermert, O2_Wormer, O2_Bussery, O2_Date}. 
As we expect that the direct magnetic interaction between NN O$_{2}$s along the $c$--axis is dominant, the same dependence can be applied. 
If so, even though the lattice dimerization in NaO$_{2}$ would be extremely so small not to be observed experimentally, 
the $J$--alternation may occur in the SP phase. 
More detailed structural study in the SP phase of NaO$_{2}$ is a future task. 

In summary, we have investigated the low--temperature magnetism of NaO$_{2}$. 
We found no magnetic phase transition down to $T=0.3$ K 
and confirmed the spin--singlet ground state below $T_{3}$. 
Raman scattering experiments clearly indicated the presence of the phase transition at $T_{3}$. 
Consequently, we conclude that the singlet--ground state of NaO$_{2}$ is due to the SP instability. 

The authors acknowledge fruitful discussions with H. O. Jescheke, J. Otsuki, M. Naka, 
K. Okada, R. Kondo, T. Goto, H. Sagayama, R. Kumai. 
The x--ray diffraction study was performed under the approval of the Photon Factory Program Advisory Committee 
(Proposal No. 2017G636, 2019T003, 2020G666). 
The neutron scattering experiment at the
HRC was approved by the Neutron Scattering Program Advisory Committee of IMSS, KEK
(proposals no. 2019S01), and ISSP. 
This work was partly supported by JSPS KAKENHI (15H03529, 20K20896, 21H04441), MEXT, Japan. 

\noindent
$^{\dag}$ Corresponding author; kambe@science.okayama-u.ac.jp\\
$^{\S}$ Present address; Institute for Molecular Science, Okazaki, Japan\\
$^{\P}$ Present address; Department of Physics, Institut Teknologi Sepuluh Nopember, Indonesia\\
$^{*}$ These authors contributed equally to this work. 


\end{document}


\title{Supplemental Material; \\Spin--gap formation due to spin--Peierls instability \\in $\pi$--orbital--ordered NaO$_2$}

\author{Mizuki Miyajima$^{1,\S, *}$, 
Fahmi Astuti$^{2,3,\P, *}$, 
Takahito Fukuda$^1$, 
Masashi Kodani$^1$, 
Shinsuke Iida$^4$, 
Shinichiro Asai$^4$, 
Akira Matsuo$^4$, 
Takatsugu Masuda$^{4,6,7}$, 
Koichi Kindo$^4$, 
Takumi Hasegawa$^5$, 
Tatsuo C Kobayashi$^{1}$, 
Takehito Nakano$^8$, 
Isao Watanabe$^{2,3}$, and 
Takashi Kambe$^{1,\dag}$ \email{kambe@science.okayama-u.ac.jp}}

\affiliation{%
$^1$Department of Physics, Okayama University, Okayama 700--8530, Japan
}%
\affiliation{%
$^2$Advanced Meson Science Laboratory, RIKEN Nishina Center, Wako, Saitama 351--0198, Japan
}%
\affiliation{%
$^3$Department of Physics, Hokkaido University, Sapporo 060--0808, Japan
}%
\affiliation{%
$^4$Institute for Solid State Physics, The University of Tokyo, Kashiwa, Chiba 277--8581, Japan
}%
\affiliation{
$^5$Graduate School of Advanced Science and Engineering, Hiroshima University, Higashi--Hiroshima, 739--8521, Japan
}
\affiliation{
$^6$ Institute of Materials Structure Science, High Energy Accelerator Research Organization, Tsukuba, Ibaraki 305--0801, Japan
}
\affiliation{
$^7$ Trans--scale Quantum Science Institute, The University of Tokyo, Tokyo 113--0033, Japan 
}
\affiliation{
$^8$Institute of Quantum Beam Science, Ibaraki University, Mito, Ibaraki 310--8512, Japan
}
\maketitle

\noindent
$^{\dag}$ Corresponding author; kambe@science.okayama-u.ac.jp\\
$^{\S}$ Present address; Institute for Molecular Science, Okazaki, Japan\\
$^{\P}$ Present address; Department of Physics, Institut Teknologi Sepuluh Nopember, Indonesia\\
$^{*}$ These authors contributed equally to this work. 

\newpage
\section{\label{sec:level1}Sample synthesis and experimental methods}
We synthesized NaO$_2$ powder samples by the solution method using liquid an ammonia NH$_3$ 
and methyl amine CH{$_3$}NH$_2$. 
In a Ar--filled glove box (O$_2$ and H$_2$O $< 0.1$ ppm), alkali metal was placed in a reaction cell 
(hyper glass cylinder produced by Taiatsu Techno Co. Ltd., ), 
and the cell was dynamically pumped down to 10$^{-2}$ Pa. 
The reaction cell was cooled with liquid N$_2$ to condense the gas mixture of NH$_3$ and CH{$_3$}NH$_2$. 
After the reaction cell was filled with liquid phase of NH$_3$ and CH{$_3$}NH$_2$ (typically, $\sim 1$ ml), 
O$_2$ gas was put in at a constant pressure of $\sim 0.1$ MPa. 
The solution was kept at $-30 ^\circ$C. 
The reaction can be recognized as complete when the solution becomes colorless and the product precipitated. 
After the reaction, we removed the liquid NH$_3$ and CH{$_3$}NH$_2$ by dynamically pumping the glass tube, 
and obtained NaO$_2$ powder. 
The color of NaO$_2$ powder is dark yellow. 
Because the NaO$_2$ sample is very sensitive to air, the samples must be handled in the Ar--filled glove box. 

The x--ray powder diffraction (xrd) patterns of the samples were measured with synchrotron radiation 
at BL--8A and 8B of KEK--PF (wave length $\lambda$ = 0.99917 \AA). 
The sample was put in a capillary. 
The He--flow and closed cycle cryostats were used for the low temperature measurements. 
Rietveld refinement was performed to obtain the structural parameters using the GSAS II package \cite{GSAS}. 
The final weighted $R$--factor, $R_{\rm wp}$, for the room temperature (RT) structure was converged to 4.36 \%, 
indicating a good fit to the experimental data. 
To refine the RT structure using the Rietveld method, we put Cl atom instead of O$_2$ molecule in the unit cell 
because of the orientational disorder of O$_2$ molecules. 
The lattice parameter of NaO$_2$ is estimated to be $a$ = 5.506 \AA, which is consistent with the literature \cite{AO2_XRD}.
Very small amount of impurity phase was found, but was not specified. 
The x--ray diffraction for the single crystal were also measured with synchrotron radiation at BL--8B of KEK--PF in order to investigate the structural distortion below $T_{3}$. 

The magnetization, $M$, was measured using a SQUID magnetometer (MPMS--R2 and MPMS3, Quantum Design Co. Ltd.) 
in the temperature region $> 2$ K. 
The low--temperature Curie--like behavior strongly depends on the sample batch. 
Although large Curie--tail easily hide the intrinsic low--temperature magnetic properties of NaO$_{2}$, 
the origin of magnetic contamination had not been specified from the xrd. 
High magnetic field magnetization was measured in pulsed magnetic fields up to 60 T below 4.2 K at ISSP. 
Van Vleck paramagnetic contribution was not taken into consideration 
because we could not estimate it either from a temperature independent term in the magnetic susceptibility experiments nor a slope above the saturation magnetization in the high magnetic field magnetization experiments. 

The muon spin relaxation ($\mu$SR) measurements were performed at the DOLLY spectrometer 
at Paul Scherrer Institut (PSI) in Switzerland and at the ARGUS spectrometer (RIKEN--RAL) in the UK. 
For $\mu$SR measurements, about 200 mg NaO$_2$ sample was packed into the plastic bag 
to prevent sample degradation. 
Zero--field (ZF) $\mu$SR measurements were carried out down to 0.3 K using Heliox--VT system (Oxford Instruments, Co. Ltd.,). 

The inelastic neutron scattering (ins) experiments were performed using high--resolution chopper spectrometer at J--PARC. 
The incident neutron energy was $30.69$ meV and the energy resolution was $2.4$ meV.
The powder sample was put in the Aluminum cell. 
The background contribution from the sample cell was measured independently 
and can be subtracted from the measurements (see section \ref{sec:level4}). 

The Raman scattering experiments were performed using a triple--axis monochrometer (JASCO, NR-1800). 
The CCD detector (Princeton Instruments, LN/CCD--1100PB) was used. 
The wave--length of the Laser was $\lambda = 564.1$ nm.  
The samples were put in the homemade Aluminum cell attached optical windows. 
The GM cryocooler (SHI, SRDK--2015) was used to cool the samples down to 4 K. 

\clearpage
\section{\label{sec:level2}Investigation of crystal structure}
\begin{figure}[H]
\begin{center}
\includegraphics[clip, width= 0.6 \textwidth]{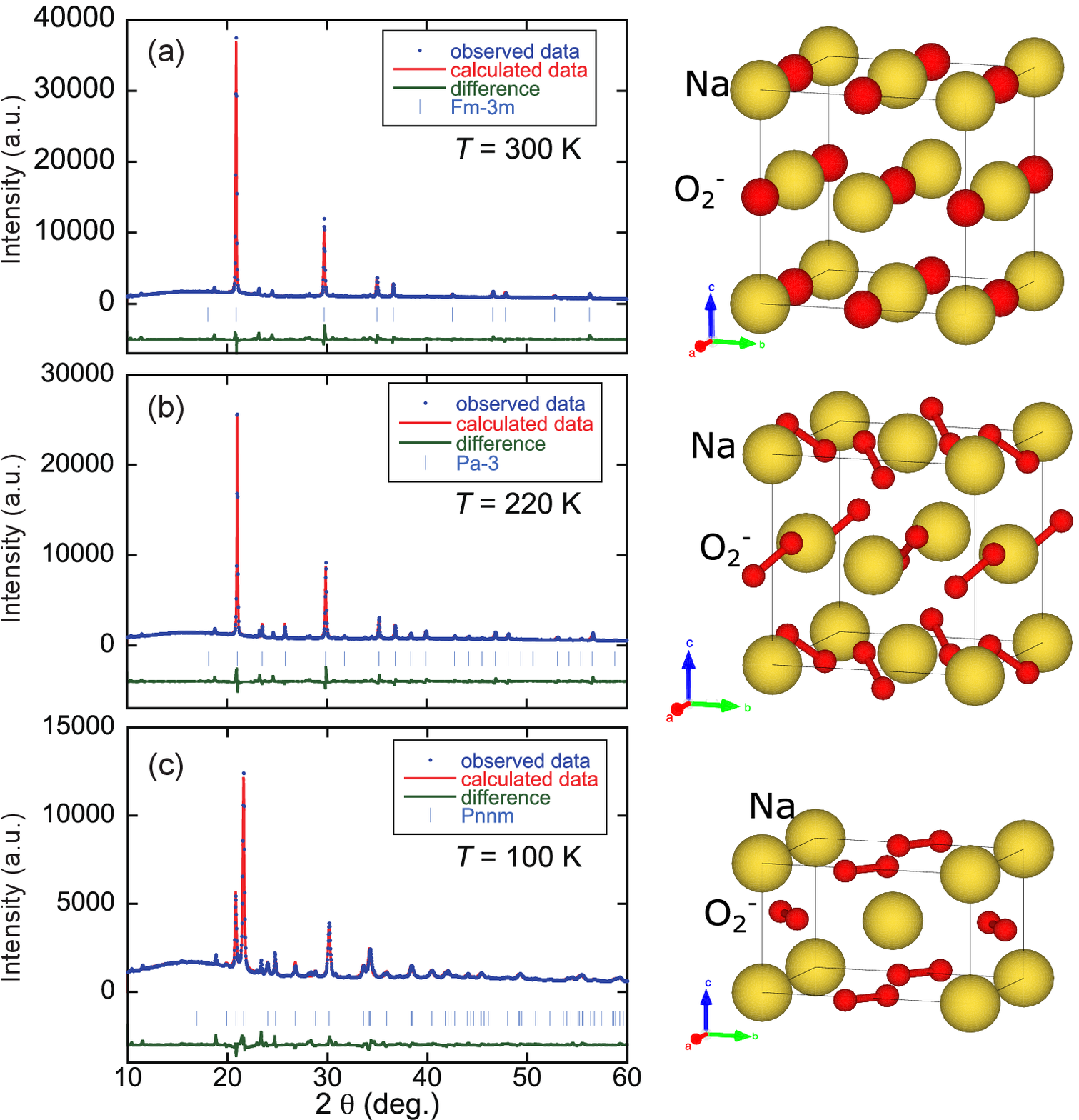}
\caption{
xrd patterns and results of Rietveld refinement at (a) RT, (b) 220 K and (c) 100 K. 
The blue dots denote the observed data, and the red line denotes the calculated results. 
The green line denotes the difference between the observed and calculated data. 
The blue vertical bars indicate the candidate position for the space group symmetry of each phase. 
The right figures show the unit cells of the phase I, II and III. 
The yellow circle and the dumbbell indicate Na atom and O$_{2}$ molecule. 
}
\label{XRPD}
\end{center}
\end{figure}

NaO$_{2}$ undergoes several successive phase transitions as discussed in the text.  
Figure \ref{XRPD} shows the xrd patterns and Rietveld refinement results 
at (a) room temperature (RT), (b) 220 K and (c) 100 K. 
All phases can be analyzed using the GSAS II package. 
The right side of Fig. \ref{XRPD} shows schematic figures of the unit cells of phase I, II and III. 
Tables \ref{table1},  \ref{table2} and \ref{table3} summarize the structural parameters at these temperatures. 

\begin{figure}[H]
\begin{center}
\includegraphics[clip, width= 0.5\textwidth]{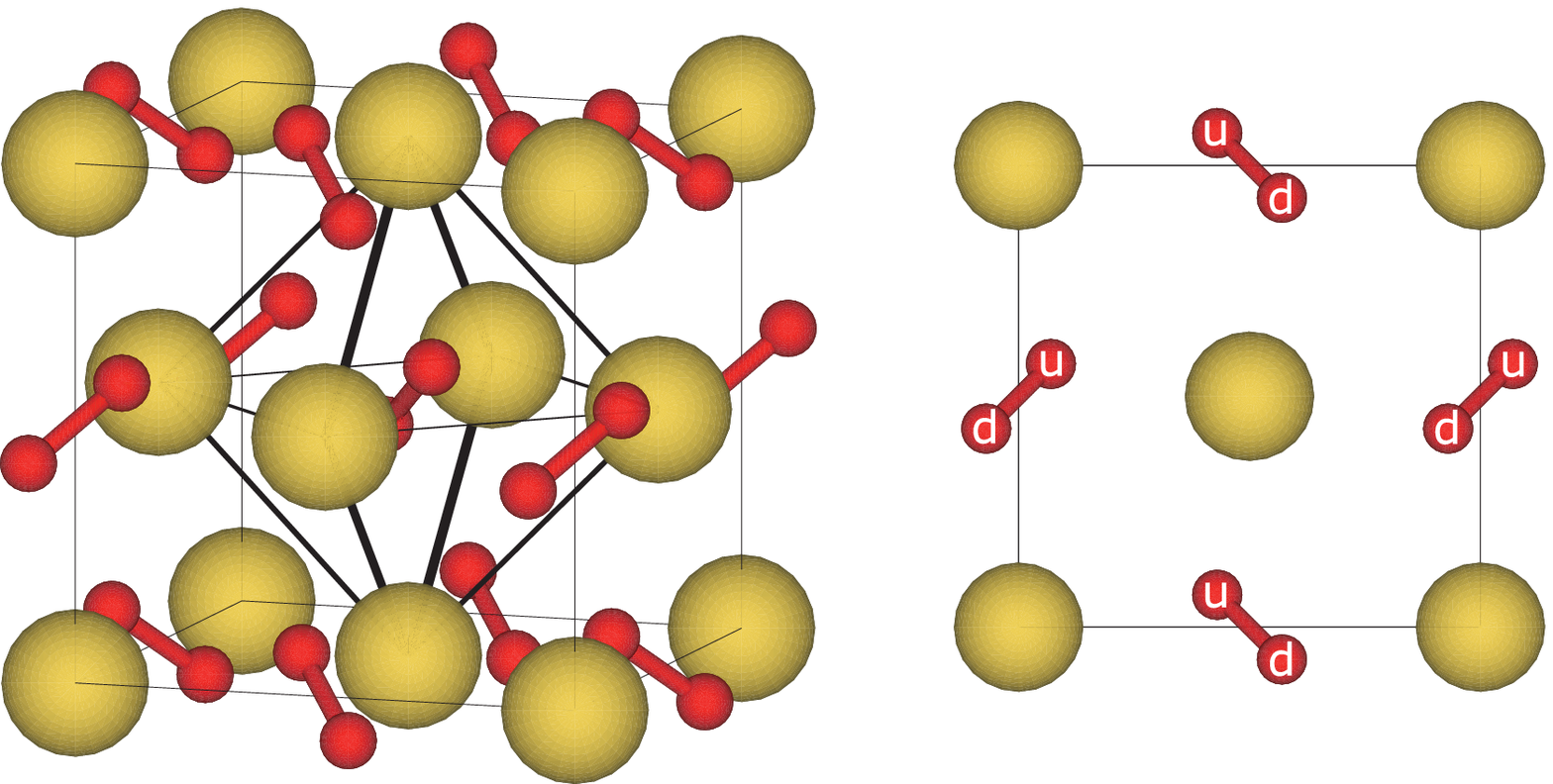}
\caption{
(left) Unit cell of the phase II of NaO$_{2}$. 
O$_{2}$ molecule is octahedrally surrounded by Na atoms.  
(right) Unit cell in the ($ac$)--plane for the phase II.
 }
\label{octahedron}
\end{center}
\end{figure}

The O$_{2}$ molecule is octahedrally surrounded by Na atoms in all phases. 
The phase I has a cubic NaCl--type structure (SG: $Fm$\=3$m$). 
Because of the cubic symmetry, O$_2$ molecular axis is not fixed along a certain direction, i.e., it has a orientational disorder. 
The phase II has a cubic pyrite--type structure (SG: $Pa$\=3). 
As shown in Fig. \ref{octahedron} (a), 
in the phase II, the O$_{2}$ molecular axis is aligned along one of four equivalent [111]--directions of the octahedron, i.e., 
[111], [11\=1], [1\=11], and [1\=1\=1]. 
Figures \ref{octahedron}(b) shows the (010) and (020) plane for the crystal structure of the phase II, 
where the {\bf u} and {\bf d} denote the upward and downward positional shift of oxygen atom against the $ac$--plane. 
In this phase, the adjacent O$_{2}$ molecules arrange their molecular axes so as to avoid each other, 
i.e., coherent {\it antiferro}--like arrangement of the molecular axes. 
The phase III has a cubic marcasite--type structure (SG: $Pnnm$). 
In the phase III, the octahedron loses the three--fold symmetry and the molecular axis is slightly 
tilted from the [111]--direction of the octahedron (see the text). 
Na--Na bond lengths are changed to be not equivalent (3.39 \AA (blue), 3.89 \AA (green) and 4.31 \AA (red)). 
Neighboring O$_{2}$ molecules within the $ac$--plane arrange their molecular axes so as to be parallel to each other, namely, 
{\it ferro}--like alignments of nearest--neighboring molecular axes are given. 

\begin{table}
\centering
\caption{Structural parameters obtained from Rietveld analysis at 300 K in the phase I. 
The weighted error $R_{\rm wp}$, Goodness Of Fitting (GOF), lattice constant $a$, atomic coordinates ($x$, $y$, $z$), 
and isotropic atomic displacement parameter $U_{\rm iso}$ are shown. 
In the Rietveld analysis, Cl atom was placed in site of the O$_{2}$ molecule. }
\begin{tabular}{ccccc} \hline
$T$ (K) & 300 & & & \\
Space Group & $Fm$\=3$m$ & & & \\
$R_{\rm wp}$ (\%) & 5.16 & & & \\
GOF & 1.77 & & & \\ \hline
$a$ (\AA) & 5.50167(7) & & & \\ \hline
atoms & $x$ & $y$ & $z$ & $U_{\rm iso}$ \\ 
Na & 0 & 0 & 0 & 0.04763\\
O$_2$ (Cl) & 0.5 & 0.5 & 0.5 & 0.27987 \\ \hline
\label{table1}
\end{tabular}
\end{table}
\begin{table}
\centering
\caption{
Structural parameters obtained from Rietveld analysis at 220 K in the phase II.  
}
\begin{tabular}{ccccc} \hline
$T$ (K) & 220 & & & \\
Space Group & $Pa$\=3 & & & \\
$R_{\rm wp}$ (\%) & 6.3 & & & \\
GOF & 2.06 & & & \\ \hline
$a$ (\AA) & 5.47875(9) & & & \\ \hline
atoms & $x$ & $y$ & $z$ & $U_{\rm iso}$ \\ 
Na & 0 & 0 & 0 & 0.01905\\
O & 0.43245 & 0.43245 & 0.43245 & 0.03352 \\ \hline
\label{table2}
\end{tabular}
\end{table}
\begin{table}
\centering
\caption{
Structural parameters obtained from Rietveld analysis at 100 K in the phase III. 
}
\begin{tabular}{ccccc} \hline
$T$ (K) & 100 & & & \\
Space Group & $Pnnm$ & & & \\
$R_{\rm wp}$ (\%) & 4.8 & & & \\
GOF & 1.49 & & & \\ \hline
$a$ (\AA) & 4.31106(45) & & & \\ 
$b$ (\AA) & 5.52452(25) & & & \\ 
$c$ (\AA) & 3.38760(25) & & & \\ \hline
atoms & $x$ & $y$ & $z$ & $U_{\rm iso}$ \\ 
Na & 0 & 0 & 0 & 0.00601\\
O & 0.11383 & 0.41468 & 0 & 0.01144 \\ \hline
\label{table3}
\end{tabular}
\end{table}
\clearpage
\section{\label{sec:level3}Investigation of magnetic exchange interaction}
\begin{figure}[H]
\begin{center}
\includegraphics[clip, width= 1 \textwidth]{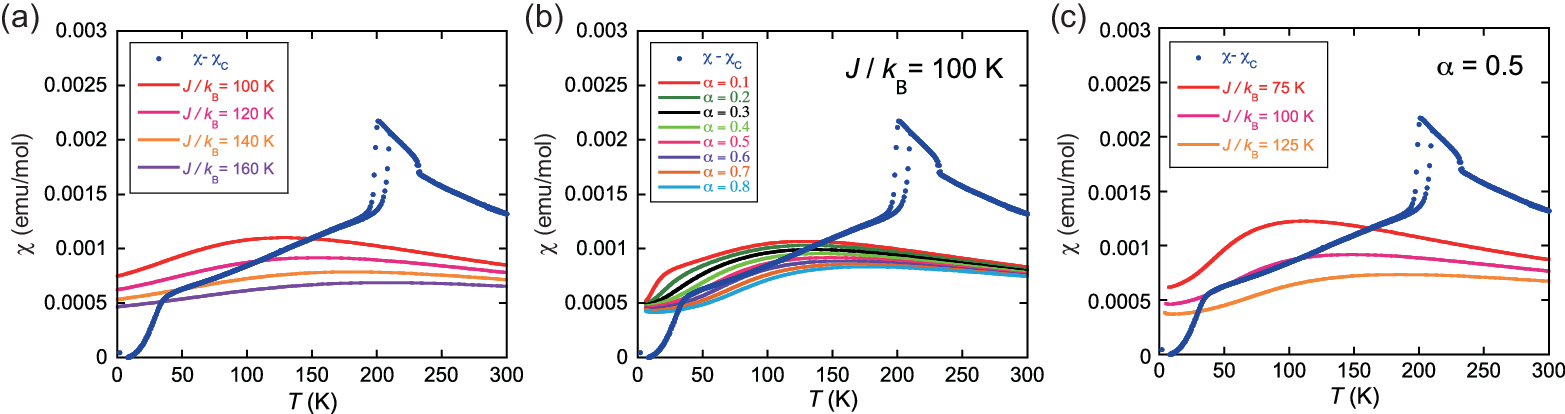}
\caption{
(a) 
Temperature dependence of magnetic susceptibility, $\chi (T)$, in powder NaO$_2$. 
Comparison $\chi (T)-\chi_{\rm C}$ with the Bonner--Fisher(BF) model, where the $\chi_{\rm C}$ 
denote the low--temperature Curie--contribution. 
Several BF curves with different $J/k_{\rm B}$ are plotted.
(b) Comparison $\chi (T)-\chi_{\rm C}$ with the two--dimensional (2DL) model with weak interchain interaction.  
The $\alpha$ denotes the ratio between intrachain, $J$, and interchain, $J'$, AF interaction. 
Several 2DL curves with different $\alpha$ are plotted when $J/k_{\rm B}=100$ K. 
(c) Comparison $\chi (T)-\chi_{\rm C}$ with the two--dimensional (2DL) model with weak interchain interaction.  
Several 2DL curves with different $J$ are plotted when $\alpha = 0.5$. 
 }
\label{J-investigation}
\end{center}
\end{figure}
Figure \ref{J-investigation} shows the temperature dependence of $\chi (T)$, in which the low--temperature Curie--contribution 
is subtracted. 
To evaluate $J$ from the $\chi (T)$ in the phase III, we used the so--called Bonner--Fisher (BF) model \cite{BF} and 
a two--dimensional (2DL) model with weak inter--chain interaction \cite{Keith}. 
In Fig. \ref{J-investigation}, $\chi (T)-\chi_{\rm C}$ and several curves for two models are plotted. 
Using the fixed Curie constant, 
we tried to fit $\chi (T)$ by these models with different exchange constants, but were unable to reproduce the experiment. 

\clearpage
\section{\label{sec:level4}Details for neutron inelastic scattering measurements}
\begin{figure}[H]
\begin{center}
\includegraphics[clip, width= 1 \textwidth]{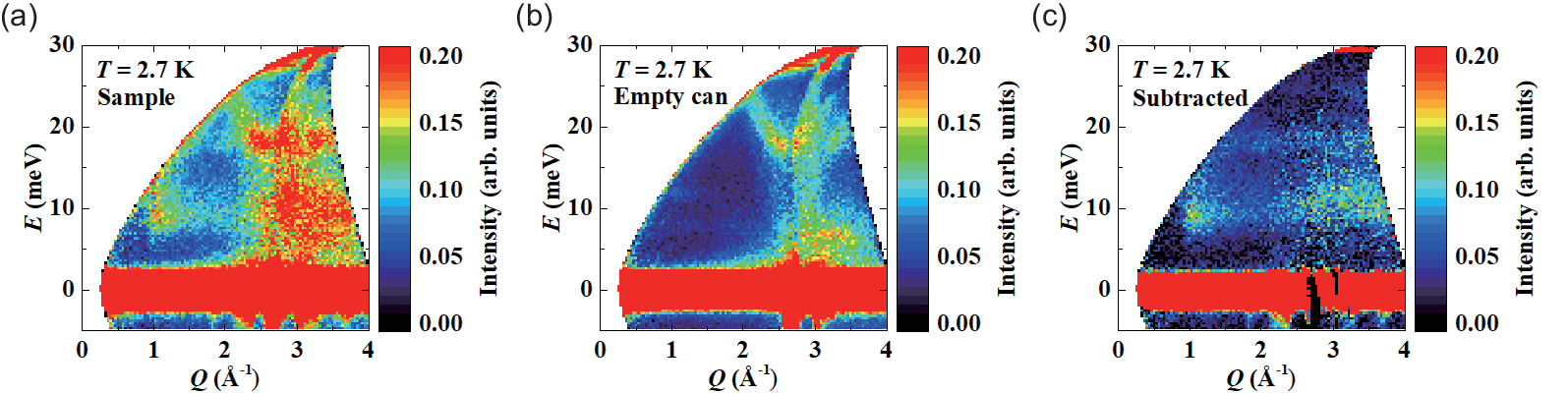}
\caption{
(a) Inelastic neutron scattering (ins) spectrum for the NaO$_{2}$ sample in an Al--cell measured 
with $E_i = 30.69$ meV at $T=2.70$ K. 
(b) ins spectrum for the Al--cell only, i.e., background spectrum (Empty can).
(c) ins spectrum subtracted (b) from (a), which corresponds to the ins spectrum for the NaO$_{2}$ sample. 
}
\label{INS1}
\end{center}
\end{figure}

\begin{figure}[H]
\begin{center}
\includegraphics[clip, width= 0.6 \textwidth]{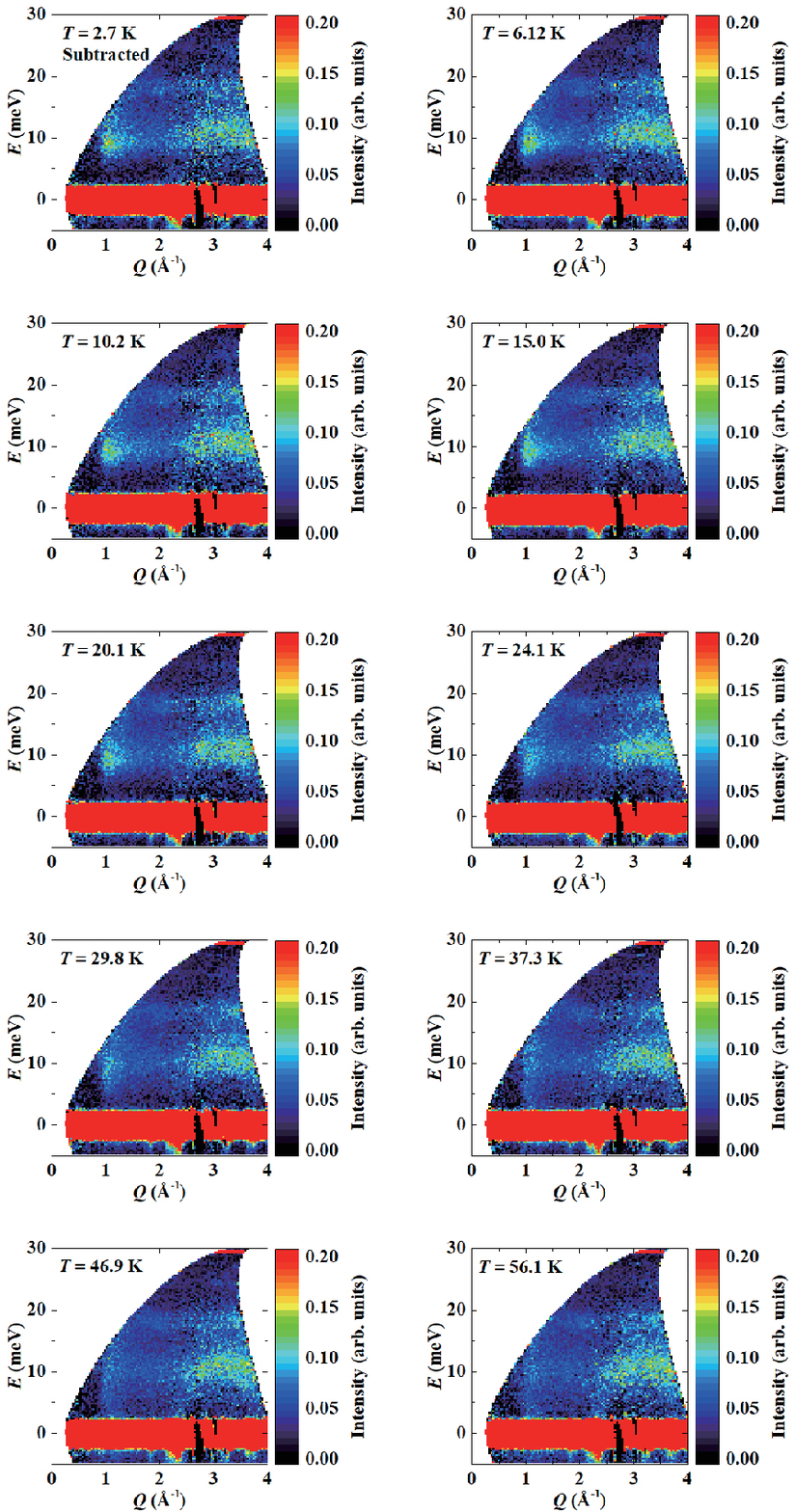}
\caption{
ins spectra for the NaO$_{2}$ sample measured at $T=$ 2.70, 6.12, 10.2, 15.0, 20.1, 24.1, 29.8, 37.3, 46.9  and 56.1 K. 
}
\label{INS2}
\end{center}
\end{figure}

Figure \ref{INS1} shows the inelastic neutron scattering (ins) spectrum measured with $E_i = 30.69$ meV at $T = 2.70$ K.
Figure \ref{INS1}(a) shows the ins spectrum for the NaO$_{2}$ sample in a sample cell made of Aluminum. 
Figure \ref{INS1} (b) shows the ins spectrum measured only for the Al--cell under the same condition, i.e., background spectrum. 
The INS spectrum obtained by subtracting (b) from (a) is shown in Fig. \ref{INS1} (c), which corresponds to 
the net spectrum for the NaO$_{2}$ sample. 
Figure \ref{INS2} shows the ins spectra for the NaO$_{2}$ sample 
at $T=$ 2.70, 6.12, 10.2, 15.0, 20.1, 24.1, 29.8, 37.3, 46.9  and 56.1 K, 
which were obtained by the same method at $T=2.70$ K. 
In order to reveal the change in the intensity of the excitation around $Q^{-1} \sim 1$ meV as a function of temperature, 
the spectra between 0.5 \AA $^{-1}$ and 1.5 \AA $^{-1}$ are integrated, which is displayed in the text. 


\clearpage
\section{\label{sec:level5}Estimation of upper limit of a distortion along the 1D chain}
The observation of the superlattice reflection in the diffraction measurements should allow us to conclude 
the SP ground state in NaO$_{2}$. 
Unfortunately, 
we could not experience the direct evidence on the structural change below $T_{3}$ 
by the powder xrd, single--crystal xrd and elastic neutron scattering measurements. 
Thus, we try to estimate the upper limit of a distortion along the $c$--axis, where the strong AF interaction would be present between the nearest--neighbor O$_2$s, by taking an experimental background systematic error into consideration in 
the elastic neutron scattering experiments. 

The simple dimerization model along the $c$--axis provides that the superlattice reflections should be observed at the index with $c^*/2$. 
The structure factor calculation leads that the intensity ratio between the (110) and (001/2) reflections is proportional 
to $(\delta c / c)^2$ if $\delta c / c \ll 1$, where the $c$ and $\delta c$ describe the length of the $c$--axis and the distortion along the $c$--axis, respectively. 
If the dimerization along the chain is assumed to be 2\%, i.e., $\delta c / c = 0.02$, the intensity ratio between the (110) and (001/2) reflection is calculated to be 0.028. 
If this is the case, we obtain the following relation; 
\begin{eqnarray}
I (001/2) &=& \frac{0.028}{(0.02)^2} \left(\frac{\delta c}{c}\right)^2 I(110) \\
dI &>& I (00 1/2)
\label{distortion}
\end{eqnarray}
From the elastic neutron scattering experiments on NaO$_2$, the integrated intensity of (110) reflection is estimated to be 7.35 (a.u.) and the background systematic error is estimated to be $dI=0.0863$ (a.u.). 
When we put these values into eq.(\ref{distortion}), the distortion, $\delta c/c$, was calculated to be less than 0.013. 
If this is the case, the distortion of the $c$--axis could not been observed.